\documentclass[10pt,prd,showpacs,showkeys]{revtex4}
\usepackage{amsmath}
\usepackage{epsfig}
\usepackage{graphicx}
\usepackage{color}
\def\NewF{
$A_4\,$}
\newcommand\B[1]{\overline{#1}}

\begin{document}

\title{
Toward minimal renormalizable SUSY $SU(5)$ Grand Unified Model \\
with tribimaximal mixing from
\NewF Flavor symmetry}

\author{Paolo Ciafaloni}
\email[]{Paolo.Ciafaloni@le.infn.it}
\author{Marco Picariello}
\email[]{Marco.Picariello@le.infn.it}
\author{Alfredo Urbano}
\email[]{ Alfredo.Urbano@le.infn.it}
\affiliation{Dip. di Fisica, Universit\`a del Salento and 
INFN - Lecce, Italy}
\author{Emilio Torrente-Lujan}
\email[]{ etl@um.es}
\affiliation{Dep. de Fisica,  Univ. de Murcia - Murcia, Spain}

\begin{abstract}
We address the problem of rationalizing the pattern of fermion masses and
mixings by adding a nonabelian flavor symmetry in a grand unified
framework. With this purpose, we include an $A_4$ flavor symmetry into a
unified renormalizable SUSY GUT SU(5) model. With the help of the ``Type II
Seesaw'' mechanism we are able to obtain the pattern of observed neutrino
mixings in a natural way, through the so called tribimaximal matrix.
\end{abstract}
\pacs{11.30.Hv, 12.10.-g, 14.60.Pq, 12.15.Ff}
\keywords{Flavor symmetries, Unified field theories and models,
 Quark and lepton masses and mixing}
\maketitle

\section{Introduction}
\noindent
The experimental discovery of flavor oscillations of neutrinos, with the consequence that their masses are different from zero,
is certainly a clear indication that there is New Physics beyond the
content of the Standard Model \cite{Altarelli:2009wt}.
One of the most attractive and beautiful scenario in which we can set this
information is represented by the Grand Unification Theories (GUT),
that describe the merging of gauge couplings into a single one at a very
high energy ($\sim 10^{16}\, \mbox{GeV}$), as
suggested 
by the gauge coupling constants running.
Inside a unification theory, moreover, it is possible also to try to 
find an answer to some important and unsolved 
questions in flavor physics: the low energy 
data described in the quark sector by the Cabibbo-Kobayashi-Maskawa mixing matrix 
as well as the hierarchy between  the quark masses.
In the leptonic sector the low energy information is far from being as
exhaustive as in the quark sector; one possibility is to assume  a particular   form for the mixing matrix: the so called 
 {\em tri-bimaximal} matrix \cite{Harrison:2002er}, which is consistent with our 
 informations coming from neutrino oscillations on neutrino mass splittings and mixing angles.
The most acclaimed possibility in order to explain the hierarchy between the masses comes from the introduction of a continuous flavor symmetry, as elegantly explained in \cite{King:2001uz}-\cite{deMedeirosVarzielas:2005qg}, while the mixing can be explained by introducing discrete symmetries.
For example, in
\cite{Mohapatra:2006pu}-\cite{Ding:2008rj}
several attempts have been done to face the flavor puzzle by introducing
discrete flavor symmetries such as $S_3$, $S_4$, $A_4$, $T'$, and so on.
%
Some attempts, as in \cite{Bazzocchi:2007au},
have been done to embed the $A_4$ flavor symmetry
into a large flavor symmetry in order to explain also the hierarchy
among the 3rd and other two generation; in particular the authors have shown that the discrete symmetry $A_4$ can help us in
solving both aspects of the flavor problem:
lepton-quark mixing hierarchy and family mass hierarchy.
The flavor symmetry $A_4$, as shown for example in \cite{Cai:2006mf,Morisi:2007ft},
is very promising also in its extension to flavor group compatible with
$SO(10)$-like grand unification. For example, by embedding
$A_4$ into a group like $SU(3)\times U(1)$, as in \cite{Bazzocchi:2008rz},
it is possible to explain both large neutrino mixing and fermion
mass hierarchy in $SO(10)$ Grand Unified Theory of Flavor (GUTF). 
Considering as underlying unification theory $SU(5)$ instead of $SO(10)$, the situation becomes very different: the Standard Model ordinary matter for each family is embedded in two distinct $SU(5)$ representations; this peculiarity makes the way in which the matter content of the theory  transforms under the action of the $A_{4}$ symmetry
not obvious, allowing for different combinations
(see for instance \cite{Altarelli:2008bg} and \cite{Ciafaloni:2009ub}).


In this paper we introduce the flavor symmetry $A_4$ in the context of a
unified $SU(5)$ theory featuring a Type II Seesaw mechanism for neutrino
masses generation. Our starting point is the Model described in
\cite{Dorsner:2007fy}, which is a renormalizable model in which no matter
fields besides the Standard Model ones are introduced. To this model we add
two ingredients: the flavor symmetry, introduced in order to produce
tribimaximal mixing in the neutrino sector, and supersymmetry, which, as we
shall see, makes the needed vacuum alignement somehow more natural.

\section{Field content and $SU(5)\otimes A_4$ invariance}\label{Higgses}
In order to clarify our notation we open now a small window on the $A_4$ proprieties, referring as an example to \cite{Altarelli:2005yx} for a more detailed discussion.
In particular in this work we use the basis where the
$A_4$ elements $S$ and $T$ acts on a $\bf 3$
multiplet as
\begin{eqnarray}\label{set}
	S=\begin{pmatrix}
	-1&0&0\\
	0&-1&0\\
	0&0&1
	\end{pmatrix}\,,
&\quad\quad&
	T=\begin{pmatrix}
	0&1&0\\
	0&0&1\\
	1&0&0
	\end{pmatrix}\,,
\end{eqnarray}
Given two triplets $(a_1,a_2,a_3)$ and  $(b_1,b_2,b_3)$, three non
equivalent singlets can be formed from the $\bf\textbf{3}\otimes\textbf{3}$
composition:

\begin{eqnarray}
	{\bf1}=a_1b_1+a_2b_2+a_3b_3\,,\quad\quad
	{\bf1}^\prime=a_1b_1+\omega^2a_2b_2+\omega a_3b_3\,,\quad\quad
	{\bf1}^{\prime\prime}=a_1b_1+\omega a_2b_2+\omega^2a_3b_3\,
\end{eqnarray}
while the two inequivalent triplets one can form are
$\{a_2b_3,a_3b_1,a_1b_2\}$ and $\{a_3b_2,a_1b_3,a_2b_1\}$. Here as usual $\omega=\exp\left(2\pi i/3\right)$. From the decomposition of the direct product $\textbf{3}\otimes \textbf{3}\otimes \textbf{3}$ we have two different singlets, as  follows:
\begin{equation}\label{duesingoletti}
(a_2 b_3 c_1+a_3b_1c_2+a_1b_2c_3),\qquad(a_3b_2c_1+a_1b_3c_2+a_2b_1c_3).
\end{equation}

We also introduce the ${\bf 4}$ representation, which is really simply a
singlet added to a triplet; this is useful in order to keep our notation
compact. For instance the Higgs multiplet belonging to a ${\bf 5}$
representation with respect to SU(5) properties, behaves 
as the direct sum  $\textbf{4}=\textbf{3}\oplus \textbf{1}$ under $A_4$:
\begin{equation}\label{clarifynotation}
\textbf{5}_{\textbf{H}}\sim \textbf{3}\oplus \textbf{1}\rightarrow \left\{\textbf{5}_{\textbf{H}}^{k=1,2,3},\,\,\widetilde{\textbf{5}}_{\textbf{H}}\right\},
\end{equation}
and one can describe the $A_4$ transformations 
in the direct sum  $\textbf{4}(\textbf{4}',\textbf{4}'')=\textbf{3}\oplus
\textbf{1}(\textbf{1}',\textbf{1}'')$ with a 4x4 matrix:
\begin{eqnarray}\label{set4}
	S_{\textbf{4},\textbf{4}',\textbf{4}''}=\begin{pmatrix}
	-1&0&0&0\\
	0&-1&0&0\\
	0&0&1&0 \\
    0&0&0&1
	\end{pmatrix}\,,
&\quad\quad&
	T_{\textbf{4},\textbf{4}',\textbf{4}''}=\begin{pmatrix}
	0&1&0&0\\
	0&0&1&0\\
	1&0&0&0\\
     0&0&0&1,\omega,\omega^{2}
	\end{pmatrix}\,.
\end{eqnarray}

We now give the SU(5) and $A_4$ field properties
we choose in this work, for Higgs (H) and
matter (T) representations, as follows:

\begin{table}[!hbt!]
\begin{tabular}{c||c|c||c|c|c|c|c|c|c}
$SU(5)$&${\bf10_\textbf{T}}$& ${\bf\B5_\textbf{T}}$&
${\bf\B5_H}$& ${\bf 5_H}$&
${\bf\B{45}_H}$&${\bf45_H}$&
${\bf\B{15}_H}$&${\bf15_H}$&
${\bf 24_H}$
\\\hline
$A_4$&$\textbf{3}$&$\textbf{3}$&$\textbf{4}$&$\textbf{4}$&$\textbf{4}$&$\textbf{4}$&$\textbf{4}''$&$\textbf{4}'$&$\textbf{1}$\\
\end{tabular}
\end{table}
\def\VsetteFields{
The field content under $SU(5)$ transformation is:
\begin{equation}\label{matterandhiggs}
\mbox{Chiral Superfields:}\,\,\,\,\,\,\,
         {\bf10_T}=\frac{1}{\sqrt{2}}\left(
\begin{array}{ccccc}
         0 & \widehat{U}_{3}^{C} & -\widehat{U}_{2}^{C} & \widehat{U}_{1} & \widehat{D}_{1} \\
         -\widehat{U}_{3}^{C} & 0 & \widehat{U}_{1}^{C} & \widehat{U}_{2} & \widehat{D}_{2} \\
         \widehat{U}_{2}^{C} & -\widehat{U}_{1}^{C} & 0 & \widehat{U}_{3} & \widehat{D}_{3} \\
         -\widehat{U}_{1} & -\widehat{U}_{2} & -\widehat{U}_{3} & 0 & \widehat{E}^{C} \\
         -\widehat{D}_{1} & -\widehat{D}_{2} & -\widehat{D}_{3} & -\widehat{E}^{C} & 0 \\
         \end{array}
         \right),\,\,\,\,\,\,
{\bf\B5_T}=\left(
\begin{array}{c}
\widehat{D}_{1}^{C} \\
\widehat{D}_{2}^{C} \\
\widehat{D}_{3}^{C} \\
\widehat{E} \\
-\widehat{N} \\
\end{array}
\right);
\end{equation}
\begin{equation}\label{HIGGSI}\mbox{Higgs Sector I:}\,\,
\left\{\begin{array}{c}
{\bf\B5_H},\, {\bf 5_H},\,{\bf\B{45}_H},\, {\bf 45_H},\,{\bf 24_H}; \\
         \\
{\bf\B{15}_H},\,{\bf15_H};
   \end{array}\right.\,\,\,\,\,\,\,\,\,\,\,\,
 \mbox{Higgs Sector II:}\,\,
\left\{\begin{array}{c}
{\bf\tilde{\B5}_H},\,{\bf\tilde5_H},\,{\bf\widetilde{\B{45}}_H},\, {\bf\widetilde{45}_H}; \\
         \\
{\bf\widetilde{\B{15}}_H},\, {\bf\widetilde{15}_H};
   \end{array}\right.
\end{equation}
}


In the Higgs sector we will introduce ${\bf 24_H}$, ${\bf\B5_H}$, ${\bf 5_H}$ in order to break spontaneously the gauge symmetry $SU(5)$ into the Standard Model one and subsequently into the residual $SU(3)_{C}\otimes U(1)_{em}$; moreover ${\bf\B{45}_H}$ and ${\bf 45_H}$ are necessary in order to avoid the wrong prediction $M_{D}^{T}=M_{E}$
while ${\bf\B{15}_H}$ and ${\bf15_H}$ will generate the right path of neutrino masses through the Higgs mechanism
implemented by the $SU(2)_{L}$ heavy scalar triplet contained into the Standard Model decomposition of $\textbf{15}_{\textbf{H}}$.

The necessity to take into account the $A_4$ assignments as explained in
the previous table is dictated by the observed phenomenology of the
masses. For instance, it is easy to show that  with 
the simpler choice of choosing 
$\textbf{5}_{\textbf{H}},\overline{\textbf{5}}_{\textbf{H}}\sim 3$ and
$\textbf{45}_{\textbf{H}},\overline{\textbf{45}}_{\textbf{H}}\sim 3$, it is 
impossible to fit the measured values for the fermion masses. Although the Higgs sector of this model could seems rather cumbersome because of the introduction of four dimensional reducible representations, we stress the fact that it rests the minimal way in which we can preserve the predictivity of the $A_{4}$ flavor symmetry in the contest of a renormalizable $SU(5)$ model.
\section{Charged fermion mass matrices}
The relevant operators in the Yukawa sector that generate the charged fermion mass matrices are
\begin{eqnarray}\label{Yukawa}
W_0&=&y_1\, {\bf10_T}\,{\bf\B5_T}\,{\bf\B5_H}
    + y_2\, {\bf10_T}\,{\bf\B5_T}\,{\bf\B{45}_H}
    + y_3\, {\bf10_T}\,{\bf10_T}\,{\bf5_H}
    + y_4\, {\bf10_T}\,{\bf10_T}\,{\bf45_H}.
\end{eqnarray}

As  will be shown in section V, in the flavor space
${\bf\B5_H},\,{\bf 5_H},\,{\bf\B{45}_H},\,{\bf 45_H}$ acquire their VEV
in the direction $\langle 1,1,1 \rangle$.  Under this condition, after
spontaneous symmetry braking  
the mass matrices
obtained from $W_0$ through (\ref{Yukawa}) are:
\begin{eqnarray}\label{eq:break1}
	M_f=\begin{pmatrix}
	h_0^f&\gamma_1^f&\gamma_2^f\\
	\gamma_2^f&h_0^f&\gamma_1^f\\
	\gamma_1^f&\gamma_2^f&h_0^f
	\end{pmatrix}
=\tilde U_\omega M_f^{\mbox{\small diag}} \tilde U_\omega^\dagger
&\quad\mbox{where}\quad &
\tilde{U}_\omega=
\frac{1}{\sqrt{3}}\left(
\begin{array}{ccc}
\omega&\omega^2&1\\
\omega^2&\omega&1\\
1&1&1
\end{array}
\right)\,,
\end{eqnarray}
where we define:
\begin{subequations}
\begin{eqnarray}
\begin{array}{rcl}
h_0^u&=&8\tilde y_3\tilde v_5\\
h_0^d&=&\tilde y_1\tilde v_{\B5}+2\tilde y_2\tilde v_{\B{45}}\\
h_0^e&=&\tilde y_1\tilde v_{\B5}-6\tilde y_2\tilde v_{\B{45}}\\
\end{array}
&\quad\quad\quad&
\begin{array}{rcl}
\gamma_{1,2}^u&=&4v_5(y_3^1+y_3^2) \pm v_{45}(y_4^1-y_4^2)\\
\gamma_{1,2}^d&=&4v_{\B5} y_1^{2,1}+2 v_{\B{45}}y_2^{2,1}\\
\gamma_{1,2}^e&=&4v_{\B5} y_1^{1,2}-6 v_{\B{45}}y_2^{1,2}
\end{array}
\end{eqnarray}
\end{subequations}
where the VEVs of the singlets from $\textbf{4}=\textbf{3}\oplus \textbf{1}$ are signed with a \emph{tilde}, $y_{i}^{j=1,2}$ refers to the two independent parameters from the singlets of $\textbf{3}\otimes \textbf{3}\otimes\textbf{3}$, as in (\ref{duesingoletti}), written for the $y_i$ Yukawa coupling in (\ref{Yukawa}) while $\widetilde{y}_{i}$ refers to the singlet from $\textbf{3}\otimes \textbf{3}\otimes \textbf{1}$. \\
Here
we notice that $h_0^f$s, $\gamma_1^f$s, and $\gamma_2^f$s are independent parameters.
The masses are given by
\begin{eqnarray}\label{eq:hier}
\begin{array}{rcl}
m_1^f&=&	|h_0^f+\gamma_1^f\omega+\gamma_2^f\omega^2|\\
m_2^f&=&	|h_0^f+\gamma_1^f\omega^2+\gamma_2^f\omega|\\
m_3^f&=&	|h_0^f+\gamma_1^f+\gamma_2^f|\\
\end{array}
\end{eqnarray}
allowing a fit of experimental values as shown in \cite{Bazzocchi:2008rz}.

As for mixing angles, since left up and down quarks have the same mass
matrix (\ref{eq:break1}), the $V_{CKM}$ is unity in first approximation. 
In order to produce the Cabibbo angle,
we now perturb the VEV directions by adding a small component in the direction
$\langle0,0,1\rangle$.
We obtain that the mass matrices are perturbed by
\begin{eqnarray}
\delta M_f=\begin{pmatrix}
0&\epsilon_1^f&0\\
\epsilon_2^f&0&0\\
0&0&0
\end{pmatrix}&\Rightarrow&
M_f^{\mbox{\small off diag}}=\tilde U_\omega^\dagger \delta M_f \tilde U_\omega=
\begin{pmatrix}
\omega \epsilon_1^f+\omega^2\epsilon_2^f&\epsilon_1^f+\epsilon_2^f&\omega^2 \epsilon_1^f+\omega\epsilon_2^f\\
\epsilon_1^f+\epsilon_2^f&\omega^2 \epsilon_1^f+\omega\epsilon_2^f&\omega \epsilon_1^f+\omega^2\epsilon_2^f\\
\omega^2 \epsilon_1^f+\omega\epsilon_2^f&\omega \epsilon_1^f+\omega^2\epsilon_2^f&\epsilon_1^f+\epsilon_2^f
\end{pmatrix}
\end{eqnarray}
The Cabibbo angle can then be generated at least in two ways:

\begin{enumerate}
\item{
As explained in \cite{Bazzocchi:2008rz} such small perturbations,
if they are of order $\lambda^5 m_3^f$ (where $\lambda$ is the
Cabibbo angle), generate the Cabibbo angle
in the quark sector and are irrelevant in the lepton sector.
The crucial point is that such assumption has the consequences that our operators give negligible
effects in the down and charged lepton sectors, since for the down and charged leptons
$M_{d,e}^{\mbox{\tiny diag}}+M_{d,e}^{\mbox{\tiny off diag}}$
remain diagonal. On the contrary for the up quarks we have that the off-diagonal entry (1,2)
cannot be neglected: the matrix $M_u^{\mbox{\tiny diag}}+M_u^{\mbox{\tiny off diag}}$ is diagonalized by a rotation
in the 12 plane with $\sin\theta_{12}\simeq\lambda$. This rotation produces the Cabibbo angle in the CKM.
}
\item{another possibility is given by assuming that the Cabibbo angle comes from a rotation in the down
sector. This can be the case if the perturbation of the $\bf\B5_{\textbf{H}}$ and $\bf\B{45}_{\textbf{H}}$, i.e. of order
$\lambda^5 m_{top}\simeq\lambda^3m_{bottom}$,
are bigger than the ones of the $\bf5_{\textbf{H}}$ and $\bf45_{\textbf{H}}$, i.e. of order $\lambda^6 m_{top}$.
Such correction generates also a small perturbation to the tri-bimaximal lepton mixing matrix of
order of the Cabibbo angle.
In particular if the dominant contribution comes from the $\bf\B5_{\textbf{H}}$ then the tri-bimaximal
lepton mixing matrix is multiplied on the left by $U_{CKM}^\dagger$ and the net result is the presence
of a non trivial quark-lepton complementarity fully compatible with the experimental data and
a prediction for the $\theta_{13}$ lepton angle \cite{Picariello:2007ss}.
On the other side, if the dominant contribution comes from the $\bf\B{45}_{\textbf{H}}$ there is a Clebsch-Gordan
coefficient between the quark and lepton mixing corrections.}
\end{enumerate}

\section{Neutrino mass matrix and lepton mixing angles}
The relevant operators that generate the neutrino mass matrix are:
\begin{eqnarray}\label{operatoredimassaneutrini}
W_1&=&\gamma\, {\bf\B5_T}\,{\bf\B5_T}\,{\bf15_H}
   +m_{\Phi}\overline{\textbf{15}}_{\textbf{H}}\textbf{15}_{\textbf{H}}.
\end{eqnarray}
We assume that the triplet from $\bf 15_{\textbf{H}}$ acquires a small VEV in the direction
$\langle0,0,1\rangle$, while we use again the \emph{tilde} for the VEV of the singlet. Under this condition the neutrino mass matrix obtained from $W_1$ is given by
\begin{eqnarray}\label{matricedimassaneutrini}
M_\nu=\begin{pmatrix}
\beta \tilde v_{15}&\gamma v_{15}&0\\
\gamma v_{15}&\omega \beta \tilde v_{15}&0\\
0&0&\omega^2 \beta \tilde v_{15}
\end{pmatrix}
=V^\star M_\nu^{\mbox{\small diag}} V^\dagger
&\quad\quad\mbox{where}\quad\quad&
V=\begin{pmatrix}
\frac{\omega}{\sqrt{2}}&0&-\frac{i\omega}{\sqrt{2}}\\
\frac{\omega^2}{\sqrt{2}}&0&\frac{i\omega^2}{\sqrt{2}}\\
0&1&0
\end{pmatrix}
\end{eqnarray}
and the lepton tri-bimaximal mixing arises:
\begin{equation}\label{eq:Vleptons}
V_{\mbox{\small leptons}}=\tilde{U}_\omega^\dagger\cdot V=
\left(
\begin{array}{ccc}
\frac{2}{\sqrt{6}} &\frac{1}{\sqrt{3}} &0 \\
-\frac{1}{\sqrt{6}} &\frac{1}{\sqrt{3}} &-\frac{1}{\sqrt{2}} \\
-\frac{1}{\sqrt{6}} &\frac{1}{\sqrt{3}} & \frac{1}{\sqrt{2}}
\end{array}
\right)\,.
\end{equation}
In (\ref{matricedimassaneutrini}) $\gamma$ is the common parameter for the two singlets from $\textbf{3}\otimes \textbf{3}\otimes\textbf{3}$, after considering that the demand for the neutrino mass matrix to be symmetric forces in (\ref{duesingoletti}) the relation $\gamma_1=\gamma_2$; $\beta$ is the parameter from the singlet of $\textbf{3}\otimes \textbf{3}\otimes \textbf{1}'$ in (\ref{operatoredimassaneutrini}).
The neutrino masses are given by
$\{|\omega^2\beta \tilde v_{15}+\gamma v_{15}|,|\omega^2\beta \tilde v_{15}|,|-\omega^2\beta \tilde v_{15}+\gamma v_{15}|\}$.\\
Since phenomenologically we have $\delta m_{12}^2>0$ we obtain
$|\beta \gamma v_{15} \tilde v_{15}|<0$ which implies $\delta m_{13}^2>0$, i.e.
a normal hierarchy; so
as a consequence the inverted hierarchy is completely ruled out in this model because of the same underlying structure imposed by the $A_4$ symmetry.\\
Finally we predict the absolute neutrino mass value and the
parameter $|m_{ee}|$ relevant for the future experiments in neutrinoless double beta decay,
 i.e.
\begin{eqnarray}
m_2\ge\frac{1}{2\sqrt{2}}
\frac{\delta m_{atm}^2+\delta m_{sol}^2}{\sqrt{\delta m_{atm}^2-\delta m_{sol}^2}}\simeq
\frac{1}{2\sqrt{2}}\sqrt{\delta m_{atm}^2}\simeq 0.02\,\, \mbox{eV}\,,
\end{eqnarray}
and
\begin{eqnarray}
|m_{ee}|\ge2 m_1+m_2\ge\frac{3}{2\sqrt{2}}\sqrt{\delta m_{atm}^2}\simeq0.05\,\,\mbox{eV}\,.
\end{eqnarray}

\section{Minimization of the potential}
The potential V is written in terms of the superpotenzial $W(\phi_i)$, 
which is an analytical function of the scalar fields $\phi_i$,
in the following way:

\begin{equation}\label{pot}
V=\sum_i|\frac{\partial W}{\partial\phi_i}|^2+V_{D-terms}+V_{soft}
\end{equation}
Here we are interested in the SU(5) and $A_4$ breaking that takes place at
scales of the order of the GUT scale; we can therefore neglect
supersymmetry breaking terms of the order of the TeV scale, described by 
$V_{soft}$: the latter  play
a crucial role in electroweak symmetry breaking, that we don't discuss. In
the following we minimize the first term in (\ref{pot}), neglecting also
D-terms: minimization then amounts  to imposing $\frac{\partial
  W}{\partial\phi_i}=0\quad\forall i$. 
After imposing this,  we show that there is a finite region in parameter space
where $V_{D-terms}=0$, justifying {\sl  a posteriori} our
assumption. 

In order to obtain a correct 
$SU(5)\to SU(3)\otimes SU(2)\otimes U(1)$ symmetry  breaking we impose
the following structure with respect to the $SU(5)$ symmetry:
\begin{equation}\label{vacuumstructure1}
\langle \textbf{45}_{\textbf{H}}\rangle^{i5}_{i}=v_{45},\,i=1,2,3;\,\,\,\,\,\,\,\,\langle \textbf{45}_{\textbf{H}}\rangle^{45}_{4}=-3v_{45};
\end{equation}
\begin{equation}\label{vacuumstructure2}
\langle\overline{\textbf{45}}_{\textbf{H}}\rangle_{i5}^{i}=v_{\overline{45}},\,i=1,2,3;\,\,\,\,\,\,\,\,
\langle\overline{\textbf{45}}_{\textbf{H}}\rangle_{45}^{4}=-3v_{\overline{45}};
\end{equation}
\begin{equation}\label{vacuum24}
\langle\textbf{24}_{H}\rangle^{\alpha}_{\,\,\alpha}=\mbox{diag}\,v_{24}^{s}\left(2v_{24}^{s},2v_{24}^{s},2v_{24}^{s},-3
v_{24}^{s}+v_{24}^{t},-3v_{24}^{s}-v_{24}^{t}\right);
\end{equation}
\begin{equation}\label{vacuum5}
\begin{array}{c}
  \langle\textbf{5}_{H}\rangle^{\alpha}=v_{5}\left(0,0,0,0,1\right)^{T}, \\
  \\
  \langle\overline{\textbf{5}}_{H}\rangle_{\alpha}=v_{\overline{5}}\left(0,0,0,0,1\right)^{T}.
\end{array}
\end{equation}
Moreover we assume that in flavor space the triplet from 
$\textbf{15}_\textbf{H}$ acquires a small vev in the direction $(0,0,1)$.

Let us now come to potential minimization. 
The renormalizable Higgs super-operators allowed under super-symmetric $SU(5)\otimes$\NewF
invariance are:
\begin{subequations}\label{superpotenziale2}
\begin{eqnarray}\label{eq:operators2}\label{eq:superpotential2}
	W_2&=&m_\Sigma\, {\bf24_{\textbf{H}}}\,{\bf24_{\textbf{H}}}
    + \lambda_\Sigma\, {\bf24_{\textbf{H}}}\,{\bf24_{\textbf{H}}}\,{\bf24_{\textbf{H}}}
    +m_5\,\overline{\textbf{5}}_{\textbf{H}}^{k}\textbf{5}_{\textbf{H}}^{k}
    + m_\Phi\,\overline{\textbf{15}}_{\textbf{H}}^{k}\textbf{15}_{\textbf{H}}^{k}
    + m_{45}\,\overline{\textbf{45}}_{\textbf{H}}^{k}\textbf{45}_{\textbf{H}}^{k}
\nonumber\\&&\quad\quad
    + \tilde{m}_5\,{\bf\widetilde{\B5}_{\textbf{H}}} {\widetilde{\bf5}_{\textbf{H}}}
    + \tilde{m}_\Phi\, {\bf\widetilde{\B{15}}_{\textbf{H}}}{\widetilde{\bf15}_{\textbf{H}}}
    + \tilde{m}_{45}\, {\bf\widetilde{\B{45}}_{\textbf{H}}}{\widetilde{\bf45}_{\textbf{H}}},
\\
W_3&=&
    \lambda_H\, \overline{\textbf{5}}_{\textbf{H}}^{k}{\bf24_{\textbf{H}}} \textbf{5}_{\textbf{H}}^{k}
    + c_H\, \overline{\textbf{5}}_{\textbf{H}}^{k} {\bf24_{\textbf{H}}} \textbf{45}_{\textbf{H}}^{k}
    + b_H\, \overline{\textbf{45}}_{\textbf{H}}^{k}{\bf24_{\textbf{H}}}\textbf{5}_{\textbf{H}}^{k}
    + a_H\, \overline{\textbf{45}}_{\textbf{H}}^{k}\textbf{45}_{\textbf{H}}^{k}{\bf24_{\textbf{H}}}
\nonumber\\&&\quad\quad
   + \tilde{\lambda}_H\, {\bf\widetilde{\B5}_{\textbf{H}}} {\bf24_{\textbf{H}}} {\widetilde{\bf5}_{\textbf{H}}}
   + \tilde{c}_H\, {\bf\widetilde{\B5}_{\textbf{H}}} {\bf24_{\textbf{H}}} {\widetilde{\bf45}_{\textbf{H}}}
    + \tilde{b}_H\, {\widetilde{\bf\B{45}}_{\textbf{H}}}{\bf24_{\textbf{H}}}{\widetilde{\bf5}_{\textbf{H}}}
    + \tilde{a}_H\, {\widetilde{\bf\B{45}}_{\textbf{H}}}{\widetilde{\bf45}_{\textbf{H}}}{\bf24_{\textbf{H}}},
\\
W_4&=&h_1 \overline{\textbf{15}}_{\textbf{H}}^{k}{\bf24_{\textbf{H}}}\textbf{15}_{\textbf{H}}^{k}
    + h_2^{lmn} \overline{\textbf{15}}_{\textbf{H}}^{l}\textbf{5}_{\textbf{H}}^{m}\textbf{5}_{\textbf{H}}^{n}
    + h_3^{lmn} \textbf{15}_{\textbf{H}}^{l}\overline{\textbf{5}}_{\textbf{H}}^{m}\overline{\textbf{5}}_{\textbf{H}}^{n}
    + h_4^{lmn} \textbf{15}_{\textbf{H}}^{l}\overline{\textbf{45}}_{\textbf{H}}^{m}\overline{\textbf{45}}_{\textbf{H}}^{n}
    + h_5^{lmn} \overline{\textbf{15}}_{\textbf{H}}^{l}\textbf{45}_{\textbf{H}}^{m}\textbf{45}_{\textbf{H}}^{n}
\nonumber\\&&\quad\quad
  + \tilde h_1 {\widetilde{\bf\B{15}}_{\textbf{H}}}\,{\bf24_{\textbf{H}}}\,{\widetilde{\bf15}_{\textbf{H}}}
    + \tilde h_2 {\bf\widetilde{\B{15}}_{\textbf{H}}}\left(\textbf{5}_{\textbf{H}}^{1}\textbf{5}_{\textbf{H}}^{1}+
    \omega^2\textbf{5}_{\textbf{H}}^{2}\textbf{5}_{\textbf{H}}^{2}+\omega
    \textbf{5}_{\textbf{H}}^{3}\textbf{5}_{\textbf{H}}^{3}
    \right)
    + \tilde h_2^\prime \overline{\textbf{15}}_{\textbf{H}}^{k}\textbf{5}_{\textbf{H}}^{k}{\widetilde{\bf5}_{\textbf{H}}}\nonumber\\&&\quad\quad
    + \tilde h_3 {\bf\widetilde{{15}}_{\textbf{H}}}\left(\overline{\textbf{5}}_{\textbf{H}}^{1}\overline{\textbf{5}}_{\textbf{H}}^{1}+
    \omega\overline{\textbf{5}}_{\textbf{H}}^{2}\overline{\textbf{5}}_{\textbf{H}}^{2}+\omega^2
    \overline{\textbf{5}}_{\textbf{H}}^{3}\overline{\textbf{5}}_{\textbf{H}}^{3}
    \right)
    + \tilde h_3^\prime \textbf{15}_{\textbf{H}}^{k}\overline{\textbf{5}}_{\textbf{H}}^{k}{\widetilde{\bf\B5}_H}
    + \tilde h_4 {\bf\widetilde{{15}}_{\textbf{H}}}\left(\overline{\textbf{45}}_{\textbf{H}}^{1}\overline{\textbf{45}}_{\textbf{H}}^{1}+
    \omega\overline{\textbf{45}}_{\textbf{H}}^{2}\overline{\textbf{45}}_{\textbf{H}}^{2}+\omega^2
    \overline{\textbf{45}}_{\textbf{H}}^{3}\overline{\textbf{45}}_{\textbf{H}}^{3}
    \right)
\nonumber\\&&\quad\quad
   + \tilde h_4^\prime \textbf{15}_{\textbf{H}}^{k}{\widetilde{\bf\B{45}}_{\textbf{H}}}\overline{\textbf{45}}_{\textbf{H}}^{k}
 + \tilde h_5 {\bf\widetilde{\B{15}}_{\textbf{H}}}\left(\textbf{45}_{\textbf{H}}^{1}\textbf{45}_{\textbf{H}}^{1}+
    \omega^2\textbf{45}_{\textbf{H}}^{2}\textbf{45}_{\textbf{H}}^{2}+\omega
    \textbf{45}_{\textbf{H}}^{3}\textbf{45}_{\textbf{H}}^{3}
    \right)
   + \tilde h_5^\prime \overline{\textbf{15}}_{\textbf{H}}^{k}{\widetilde{\bf{45}}_{\textbf{H}}}\textbf{45}_{\textbf{H}}^{k},
\end{eqnarray}
\end{subequations}
where $\gamma$, $\beta$, $a_H$, $b_H$, and $c_H$
and the $y$s, $\lambda$s, $m$s, and $h$s are the coupling constants of the
model.
The invariant combination from $\textbf{3}\otimes \textbf{3}\otimes
\textbf{3}$, e.g. as abbreviated in $h_2^{lmn}
\overline{\textbf{15}}_{\textbf{H}}^{l}\textbf{5}_{\textbf{H}}^{m}\textbf{5}_{\textbf{H}}^{n}$,
have to be understood following (\ref{duesingoletti}).

We now impose  $\frac{\partial
  W}{\partial\phi_i}=0\forall i$, the superpotenzial W being given by the
  sum of the terms (\ref{superpotenziale2}a-\ref{superpotenziale2}c). The
  first equations we discuss are the ones obtained by imposing
 $\partial W/\partial \textbf{45}_{\textbf{H}}^{k}=\partial
  W/\partial \overline{\textbf{45}}_{\textbf{H}}^{k}=0$:

\begin{subequations}\label{primaderivata}
\begin{eqnarray}
  v_{\overline{45}}^{k}A &=& -c_{H}v_{24}^{s}v_{\overline{5}}^{k}, \\
  3v_{\overline{45}}^{k}B &=& -\frac{c_{H}}{2}(-3v_{24}^{s}+v_{24}^{t})v_{\overline{5}}^{k}, \\
  v_{45}^{k}A &=& -b_{H}v_{24}^{s}v_{5}^{k}, \\
 3 v_{45}^{k}B &=& -\frac{b_{H}}{2}(-3v_{24}^{s}+v_{24}^{t})v_{5}^{k};
\end{eqnarray}\end{subequations}
where
\begin{subequations}\label{AB}
\begin{eqnarray}
 A&\equiv& m_{45}+v_{24}^{s}\left(2a_{H}^{1}+\frac{a_{H}^{2}}{2}\right)+v_{24}^{t}\frac{a_{H}^{2}}{2}, \\
B&\equiv& -m_{45}+3v_{24}^{s}(a_{H}^{1}-a_{H}^{2})-v_{24}^{t}a_{H}^{1}
\end{eqnarray}\end{subequations}
Eqs.  (\ref{primaderivata}) imply that $v_5(v_{\overline{5}})$ is aligned 
with $v_{45}(v_{\overline{45}})$ in flavor space.

 Then, from $\partial W/\partial \widetilde{\textbf{15}}_{\textbf{H}}
=\partial W/\partial \widetilde{\overline{\textbf{15}}}_{\textbf{H}}=0$ we obtain:
\begin{subequations}\label{sestaderivata}
\begin{eqnarray}
  12\widetilde{h}_{4}\left[(v_{\overline{45}}^{1})^{2}+\omega(v_{\overline{45}}^{2})^{2}+\omega^{2}(v_{\overline{45}}^{3})^{2}\right]+
  \widetilde{h}_{3}\left[(v_{\overline{5}}^{1})^{2}+\omega(v_{\overline{5}}^{2})^{2}+\omega^{2}(v_{\overline{5}}^{3})^{2}\right] &=& 0 \\
  12\widetilde{h}_{5}\left[(v_{45}^{1})^{2}+\omega(v_{45}^{2})^{2}+\omega^{2}(v_{45}^{3})^{2}\right]+
  \widetilde{h}_{2}\left[(v_{5}^{1})^{2}+\omega(v_{5}^{2})^{2}+\omega^{2}(v_{5}^{3})^{2}\right] &=& 0
\end{eqnarray}
\end{subequations} 
For generic values of the superpotential parameters $\widetilde{h}_i$, these
equations are identically satisfied (recall that
  $\omega=\exp[\frac{2\pi i}{3}]$) if  
 $v_5^1=v_5^2=v_5^3$ and the same holds for $v_{45}^{i},
v_{\overline{45}}^{i}, v_{\overline{5}}^{i}$: this realizes the desired
vacuum alignment since all triplets VEVs must be proportional to the
direction (1,1,1) in flavor space.

Let us now consider the remaining equations, with the purpose of showing
that a nontrivial solution indeed exists, provided certain conditions are
fulfilled by the parameters of the superpotential.

$\ast)$\,\,from $\partial W/\partial \textbf{5}_{\textbf{H}}^{k}=0$ and
$\partial W/\partial \overline{\textbf{5}}_{\textbf{H}}^{k}=0$ we obtain:\\
\begin{subequations}\label{secondaderivata}
\begin{eqnarray}
  v_{5}^{k}\alpha &=& 3c_{H}v_{45}^{k}(-5v_{24}^{s}+v_{24}^{t}), \\
  v_{\overline{5}}^{k}\alpha &=& 3b_{H}v_{\overline{45}}^{k}(-5v_{24}^{s}+v_{24}^{t});
\end{eqnarray}
\end{subequations}

$\ast)$\,\,from $\partial W/\partial \widetilde{\textbf{45}}_{H}=0$ and $\partial W/\partial \widetilde{\overline{\textbf{45}}}_{\textbf{H}}=0$:\\
\begin{subequations}\label{terzaderivata}
\begin{eqnarray}
  \widetilde{v}_{\overline{45}}\widetilde{A} &=& -\widetilde{c}_{H}v_{24}^{s}\widetilde{v}_{\overline{5}}, \\
  3\widetilde{v}_{\overline{45}}\widetilde{B} &=& -\frac{\widetilde{c}_{H}}{2}(-3v_{24}^{s}+v_{24}^{t})\widetilde{v}_{\overline{5}}, \\
  \widetilde{v}_{45}\widetilde{A} &=& -\widetilde{b}_{H}v_{24}^{s}v_{5}, \\
 3 \widetilde{v}_{45}\widetilde{B} &=& -\frac{\widetilde{b}_{H}}{2}(-3v_{24}^{s}+v_{24}^{t})\widetilde{v}_{5};
\end{eqnarray}
\end{subequations}

$\ast)$\,\,from $\partial W/\partial \widetilde{\textbf{5}}_{\textbf{H}}=0$ and $\partial W/\partial \widetilde{\overline{\textbf{5}}}_{\textbf{H}}=0$:\\
\begin{subequations}\label{quartaderivata}
\begin{eqnarray}
  \widetilde{v}_{5}\widetilde{\alpha} &=& 3\widetilde{c}_{H}\widetilde{v}_{45}(-5v_{24}^{s}+v_{24}^{t}), \\
  \widetilde{v}_{\overline{5}}\widetilde{\alpha} &=& 3\widetilde{b}_{H}\widetilde{v}_{\overline{45}}(-5v_{24}^{s}+v_{24}^{t});
\end{eqnarray}
\end{subequations}

$\ast)$\,\,from $\partial W/\partial \textbf{15}_{\textbf{H}}^{k}=0$ and
$\partial W/\partial \overline{\textbf{15}}_{\textbf{H}}^{k}=0$ 
 for every  $l\neq m\neq k$ we obtain:\\
\begin{subequations}\label{quintaderivata}
\begin{eqnarray}
  12(h_{4}^1+h_4^2)v_{\overline{45}}^{l}v_{\overline{45}}^{m}+12\widetilde{h}_4^{\prime}\widetilde{v}_{\overline{45}}v_{\overline{45}}^{k}+(h_{3}^{1}+h_{3}^{2})
  v_{\overline{5}}^{l}v_{\overline{5}}^{m}+\widetilde{h}_{3}^{\prime}\widetilde{v}_{\overline{5}}v_{\overline{5}}^{k} &=& 0, \\
  12(h_{5}^1+h_5^2)v_{45}^{l}v_{45}^{m}+12\widetilde{h}_5^{\prime}\widetilde{v}_{45}v_{45}^{k}+(h_{2}^{1}+h_{2}^{2})
  v_{5}^{l}v_{5}^{m}+\widetilde{h}_{2}^{\prime}\widetilde{v}_{5}v_{5}^{k} &=& 0,
\end{eqnarray}
\end{subequations}

$\ast)$\,\,from $\partial W/\partial \textbf{24}_{\textbf{H}}=0$:\\
\begin{equation}\label{prima24}
  2v_{24}^{s}\beta_{1}+\sum_{k=1}^{3}\left[(2a_{H}^{1}-a_{H}^{2})v_{\overline{45}}^{k}
  v_{45}^{k}+b_{H}v_{\overline{45}}^{k}v_{5}^{k}+c_{H}v_{45}^{k}v_{\overline{5}}^{k}\right]+(2\widetilde{a}_{H}^{1}-\widetilde{a}_{H}^{2})
  \widetilde{v}_{\overline{45}}\widetilde{v}_{45}+\widetilde{b}_{H}\widetilde{v}_
  {\overline{45}}\widetilde{v}_{5}+\widetilde{c}_{H}\widetilde{v}_{45}\widetilde{v}_{\overline{5}}=0
\end{equation}
\begin{align}\label{seconda24}
  &(-3v_{24}^{s}+v_{24}^{t})\beta_{2}+\nonumber\\
  &3\left\{\sum_{k=1}^{3}\left[3(2a_{H}^{1}-a_{H}^{2})v_{\overline{45}}^{k}
  v_{45}^{k}-b_{H}v_{\overline{45}}^{k}v_{5}^{k}-c_{H}v_{45}^{k}v_{\overline{5}}^{k}\right]+3(2\widetilde{a}_{H}^{1}-\widetilde{a}_{H}^{2})
  \widetilde{v}_{\overline{45}}\widetilde{v}_{45}-\widetilde{b}_{H}\widetilde{v}_
  {\overline{45}}\widetilde{v}_{5}-\widetilde{c}_{H}\widetilde{v}_{45}\widetilde{v}_{\overline{5}}\right\}=0
\end{align}
\begin{equation}\label{terza24}
(-3v_{24}^{s}-v_{24}^{t})\beta_{3}+\sum_{k=1}^{3}\left[\lambda_{H}v_{\overline{5}}^{k}v_{5}^{k}-12a_{H}^{2}v_{\overline{45}}^{k}v_{45}^{k}\right]+
\widetilde{\lambda}_{H}\widetilde{v}_{\overline{5}}\widetilde{v}_{5}-12\widetilde{a}_{H}^{2}\widetilde{v}_{\overline{45}}\widetilde{v}_{45}=0
\end{equation}
where we have defined the following combinations:
\begin{subequations}
\begin{eqnarray}\label{valorisenzatilde}
\alpha &\equiv& m_{5}-\lambda_{H}(3v_{24}^{s}+v_{24}^{t}),\\
\beta_1&\equiv& (2m_{\Sigma}+6\lambda_{\Sigma}v_{24}^{s}),\\
\beta_{2}&\equiv& \left[2m_{\Sigma}+3\lambda_{\Sigma}(-3v_{24}^{s}+v_{24}^{t})\right],\\
\beta_{3}&\equiv& \left[2m_{\Sigma}+3\lambda_{\Sigma}(-3v_{24}^{s}-v_{24}^{t})\right],
\end{eqnarray}
\end{subequations}
with similar relations for $\widetilde{A}$, $\widetilde{B}$ (see
 eqs. \ref{AB})
 and $\widetilde{\alpha}$, obtained considering the substitutions of the ``non-tilded" parameters with the ``tilded" ones.

Comparing the first equation in (\ref{primaderivata}) with the second one, as well as the third with the fourth, and performing the same analysis with (\ref{terzaderivata}), we obtain the relations:
\begin{equation}\label{primacondizione}
\left\{\begin{array}{c}
         6Bv_{24}^{s}=A(-3v_{24}^{s}+v_{24}^{t}) \\
         6\widetilde{B}v_{24}^{s}=\widetilde{A}(-3v_{24}^{s}+v_{24}^{t})
       \end{array}
\right.\rightarrow \frac{B}{A}=\frac{\widetilde{B}}{\widetilde{A}};
\end{equation}
from (\ref{primaderivata}) and (\ref{secondaderivata}) we have, instead:
\begin{equation}\label{secondacondizione}
\left\{\begin{array}{c}
         3b_{H}c_{H}v_{24}^{s}(-5v_{24}^{s}+v_{24}^{t})=-\alpha A \\
         3\widetilde{b}_{H}\widetilde{c}_{H}v_{24}^{s}(-5v_{24}^{s}+v_{24}^{t})=-\widetilde{\alpha} \widetilde{A}
       \end{array}
\right.\rightarrow \frac{\alpha A}{b_{H}c_{H}}=\frac{\widetilde{\alpha} \widetilde{A}}{\widetilde{b}_{H}\widetilde{c}_{H}};
\end{equation}
it's possible, at this point, to use the system of (\ref{primacondizione},\ref{secondacondizione}) in order to obtain $v_{24}^{s,t}$ as functions of the parameters in the superpotential. The allowed solutions are:
\begin{equation}\label{soluzione}
v_{24}^{t}=\frac{3\eta\pm\sqrt{2\varphi}}{4\sigma},\,\,\,\,\,\,\,\,\,\,\,\,\,\,v_{24}^{s}=\frac{\eta\mp\sqrt{2\varphi}}{12\sigma},
\end{equation}
where:
\begin{subequations}
\begin{eqnarray}
  \eta &\equiv& (2a_{H}^{1}-a_{H}^{2})m_{5}b_{H}c_{H}+(2a_{H}^{1}-a_{H}^{2})^{2}m_{5}\lambda_{H}, \\
  \sigma &\equiv& \left[b_{H}c_{H}+\lambda_{H}(2a_{H}^{1}-a_{H}^{2})\right]^{2}, \\
  \varphi &\equiv& m_{5}\sigma\left[3m_{45}b_{H}c_{H}+(2a_{H}^{1}-a_{H}^{2})\left(3m_{45}\lambda_{H}+m_{5}(a_{H}^{1}+a_{H}^{2})\right)\right].
\end{eqnarray}
\end{subequations}

From  (\ref{quintaderivata})  we  obtain:
\begin{subequations}\label{relazionitildate1}
\begin{eqnarray}
  \widetilde{v}_{5} &=& -\frac{\widetilde{A}}{A}\left[\frac{12(h_{5}^{1}+h_{5}^{2})(b_{H}v_{24}^{s})^{2}+A^{2}(h_{2}^{1}+h_{2}^{2})}
  {12\widetilde{h}_{5}^{\prime}b_{H}\widetilde{b}_{H}(v_{24}^{s})^{2}+\widetilde{h}_{2}^{\prime}A\widetilde{A}}\right]v_{5}\equiv -\frac{\widetilde{A}}{A} \Theta_{1}v_{5}, \\
   \widetilde{v}_{\overline{5}} &=& -\frac{\widetilde{A}}{A}\left[\frac{12(h_{4}^{1}+h_{4}^{2})(c_{H}v_{24}^{s})^{2}+A^{2}(h_{3}^{1}+h_{3}^{2})}
  {12\widetilde{h}_{4}^{\prime}c_{H}\widetilde{c}_{H}(v_{24}^{s})^{2}+\widetilde{h}_{3}^{\prime}A\widetilde{A}}\right]v_{\overline{5}}\equiv -\frac{\widetilde{A}}{A}\Theta_{2}v_{\overline{5}}
 \end{eqnarray}
\end{subequations}
and:
\begin{subequations}\label{relazionitildate2}
\begin{eqnarray}
  \widetilde{v}_{45} &=& -\frac{\widetilde{b}_{H}v_{24}^{s}}{A}\Theta_{1}v_{5}, \\
  \widetilde{v}_{\overline{45}} &=& -\frac{\widetilde{c}_{H}v_{24}^{s}}{A}\Theta_{2}v_{\overline{5}};
\end{eqnarray}
\end{subequations}
 These relations allow us to express $v_{45}$, $\widetilde{v}_{45}$ and $\widetilde{v}_{5}$ as functions of $v_{5}$, as well as $v_{\overline{45}}$, $\widetilde{v}_{\overline{45}}$ and $\widetilde{v}_{\overline{5}}$ as functions of $v_{\overline{5}}$.\\
Considering the relations obtained in
(\ref{relazionitildate1},\ref{relazionitildate2}), we can rewrite the three
equations from (\ref{prima24},\ref{seconda24},\ref{terza24}) as three
compatible relations that allow to write
 the product $v_{5}\,v_{\overline{5}}$ as
a function of the parameters of the superpotential. 

\def\doublettriplet{
\section{Doublet-Triplet Splitting}

As usual in a SUSY GUT context based on $SU(5)$ gauge symmetry, our model suffers of the doublet-triplet splitting problem.
 Since in this work we are interested
about phenomenological consequences arising from the introduction of a flavor symmetry, we consider just the usual fine tuning approach
to the solution of this
problem, even if most interesting proposals have recently been done, leaving us with the possibility of a larger reexamination of the entire model.\\
In the following we consider however the Higgs doublet-triplet supersymmetric content of the theory, showing which one parameters have to be fine tuned in order to obtain the correct path of spontaneous symmetry breaking. We use the following decomposition of $\textbf{5}_{H}$ and $\overline{\textbf{5}}_{H}$ supermultiplets:
\begin{equation}\label{supermultipletti5}
  \textbf{5}_{H}=\left(
                   \begin{array}{c}
                     \widehat{T}_{5} \\
                     \widehat{H}_{5} \\
                   \end{array}
                 \right),\,\,\,\,\,\,\,\,\,\,\,\overline{\textbf{5}}_{H}=\left(
                   \begin{array}{c}
                     \widehat{T}_{\overline{5}} \\
                     \widehat{H}_{\overline{5}} \\
                   \end{array}
                 \right),
\end{equation}
where as usual:
\begin{equation}\label{componenteneutra}
  \widehat{H}_{5}=\left(
                      \begin{array}{c}
                        \widehat{H}_{5}^{+} \\
                        \widehat{H}_{5}^{0} \\
                      \end{array}
                    \right),\,\,\,\,\,\,\,\,\,\,\,\widehat{H}_{\overline{5}}=\left(
                      \begin{array}{c}
                        \widehat{H}_{\overline{5}}^{-} \\
                        \widehat{H}_{\overline{5}}^{0} \\
                      \end{array}
                    \right);
\end{equation}
considering the superpotential in (\ref{superpotenziale}), the only relevant terms respect to the doublet-triplet splitting problem are:
\begin{equation}\label{rilevanti}
W_{D/T}=m_5\,{\bf\B5_H} {\bf5_H}+\lambda_H\, {\bf\B5_H} {\bf24_H} {\bf5_H}
    + c_H\, {\bf\B5_H} {\bf24_H} {\bf45_H}
    + b_H\, {\bf5_H}{\bf24_H}{\bf\B{45}_H}
\end{equation}
with similar relation for the ``tilded" case. The only relevant difference of our approach compared to the usual one comes from the presence of the interaction terms; in fact, writing from the superpotential the scalar potential and considering the first step of spontaneous symmetry breaking at GUT scale, that is $\langle \textbf{24}_{H}\rangle=v_{24}^{s}\mbox{diag}(2,2,2,-3,-3)$, we obtain the quadratic terms:
\begin{subequations}\label{terminiquadratici}
\begin{equation}
\left\{|m_{5}-3\lambda_{H}v_{24}^{s}|^{2}+21|b_{H}v_{24}^{s}|^{2}\right\}H^{\dag}_{5}H_{5}+
\left\{|m_{5}+2\lambda_{H}v_{24}^{s}|^{2}+26|b_{H}v_{24}^{s}|^{2}\right\}T^{\dag}_{5}T_{5},
\end{equation}
\begin{equation}
\left\{|m_{5}-3\lambda_{H}v_{24}^{s}|^{2}+21|c_{H}v_{24}^{s}|^{2}\right\}H^{\dag}_{\overline{5}}H_{\overline{5}}+
\left\{|m_{5}+2\lambda_{H}v_{24}^{s}|^{2}+26|c_{H}v_{24}^{s}|^{2}\right\}T^{\dag}_{\overline{5}}T_{\overline{5}},
\end{equation}
\end{subequations}
with similar relations for the ``tilded" case. As we can see in this case the fine tuning involves also the parameters $c_{H}$ and $b_{H}$.\\
}

We now show that it is possible to choose the (super)potential parameters
in such a way that the D-terms contribution appearing in (\ref{pot}) are
zero. For a supersymmetric gauge theory the D-terms can be written as:
\begin{equation}\label{dterms}
\frac{1}{2}\sum_{G}\sum_{\alpha}\sum_{i,j}g_{G}^{2}\left(\phi_{i}^{\dag}T_{G}^{\alpha}\phi_{i}\right)
\left(\phi_{j}^{\dag}T_{G}^{\alpha}\phi_{j}\right),
\end{equation}
where we take into account that, for the  MSSM, 
$G=SU(3)_{C},\,SU(2)_{L},\,U(1)_{Y}$, 
with different couplings $g_{G}$ and generators $T_{G}$.

Let us first consider contributions for 
$\textbf{5}_{\textbf{H}},\overline{\textbf{5}}_{\textbf{H}}$
representations. The following decomposion holds:
\begin{equation}
  \textbf{5}_{\textbf{H}} = (\textbf{3},\textbf{1},-\textbf{1}/\textbf{3})\oplus (\textbf{1},\textbf{2},\textbf{1}/\textbf{2}),\qquad
  \overline{\textbf{5}}_{\textbf{H}} =
  (\overline{\textbf{3}},\textbf{1},\textbf{1}/\textbf{3})\oplus (\textbf{1},\overline{\textbf{2}},-\textbf{1}/\textbf{2})\label{silvestri};
\end{equation}
Since we only consider contributions to D-terms coming from the vevs 
$\langle
\textbf{5}_{\textbf{H}}\rangle,\langle
\overline{\textbf{5}}_{\textbf{H}}\rangle$, only the $SU(2)\otimes U(1)$
doublet in (\ref{silvestri})
contributes. Moreover the off diagonal SU(2) generators 
$T_1,T_2$ also give zero contribution, so we need only to consider the
effect of $T_3$ and the hypercharge $Y$. Taking also into account that in
flavor space the vevs have the structure  $\langle
\textbf{5}_{\textbf{H}},\overline{\textbf{5}}_{\textbf{H}}\rangle
=v_{5,\overline{5}}(1,1,1)$, a straightforward calculation gives:
 \begin{equation}\label{kio}
    \langle\,\,\left|\textbf{5}_{\textbf{H}}^{\dag}T_{5}^{\alpha}\textbf{5}_{\textbf{H}}\right|_{SU(2)_{L}}+
\left|\overline{\textbf{5}}_{\textbf{H}}^{\dag}T_{\overline{5}}^{\alpha}\overline{\textbf{5}}_{\textbf{H}}\right|_{SU(2)_{L}}\,\,\rangle=
\frac{3}{2}\left(-|v_{5}|^{2}+|v_{\overline{5}}|^{2}\right)+\frac{1}{2}\left(-|\widetilde{v}_{5}|^{2}+|\widetilde{v}_{\overline{5}}|^{2}\right)
\end{equation}
while the U(1) contribution reads:
 \begin{equation}\label{examplevacuum}
\langle\,\,\left|\textbf{5}_{\textbf{H}}^{\dag}T_{5}^{\alpha}\textbf{5}_{\textbf{H}}\right|_{U(1)_{Y}}+
\left|\overline{\textbf{5}}_{\textbf{H}}^{\dag}T_{\overline{5}}^{\alpha}\overline{\textbf{5}}_{\textbf{H}}\right|_{U(1)_{Y}}\,\,\rangle=
\frac{3}{2}\left(|v_{5}|^{2}-|v_{\overline{5}}|^{2}\right)+
\frac{1}{2}\left(|\widetilde{v}_{5}|^{2}-|\widetilde{v}_{\overline{5}}|^{2}\right).
\end{equation}
Similar considerations hold for the
$\textbf{45}_{\textbf{H}},\overline{\textbf{45}}_{\textbf{H}}$
representations, decomposed as
\begin{align}
\textbf{45}_{\textbf{H}}=&(\textbf{8},\textbf{2},\textbf{1}/\textbf{2})
\otimes(\overline{\textbf{6}},\textbf{1},-\textbf{1}/\textbf{3})\otimes(\textbf{3},\textbf{3},-\textbf{1}/\textbf{3})
\otimes(\overline{\textbf{3}},\textbf{2},-\textbf{7}/\textbf{6})\otimes(\textbf{3},\textbf{1},-\textbf{1}/\textbf{3})\nonumber\\
&\oplus(\overline{\textbf{3}},\textbf{1},\textbf{4}/\textbf{3})\otimes(\textbf{1},\textbf{2},\textbf{1}/\textbf{2})
\end{align}
and for which only the doublet component contributes. The
$\textbf{24}_{\textbf{H}}$ instead:
\begin{equation}\label{24}
\textbf{24}_{\textbf{H}}=(\textbf{8},\textbf{1},\textbf{0})\oplus(\textbf{1},\textbf{3},\textbf{0})
\oplus(\textbf{3},\textbf{2},-\textbf{5}/\textbf{6})\oplus(\overline{\textbf{3}},\overline{\textbf{2}},\textbf{5}/\textbf{6})
\oplus(\textbf{1},\textbf{1},\textbf{0})
\end{equation}
acquires a nonzero vev along the $(\textbf{1},\textbf{1},\textbf{0})$ component,
which is an isospin singlet with zero hypercharge and therefore does not
contribute to D-terms. Overall, D-terms can be written as:
\begin{equation}\label{dtermsprimo}
\frac{g^{2}+g'^2}{2}\left\{\left[\frac{3}{2}\left(-|v_{5}|^{2}+|v_{\overline{5}}|^{2}\right)+
\frac{1}{2}\left(-|\widetilde{v}_{5}|^{2}+|\widetilde{v}_{\overline{5}}|^{2}\right)+
\frac{3}{2}\left(-|v_{45}|^{2}+|v_{\overline{45}}|^{2}\right)+
\frac{1}{2}\left(-|\widetilde{v}_{45}|^{2}+|\widetilde{v}_{\overline{45}}|^{2}\right)
\right]\right\}^2
\end{equation}

Since all vevs appearing in (\ref{dtermsprimo}) are expressed as functions
of $v_{5}$ and $v_{\overline{5}}$ through equations
(\ref{primaderivata},\ref{relazionitildate1},
\ref{relazionitildate2}), imposing vanishing
D-terms implies:
\begin{equation}\label{FEWF}
|v_{5}|^2\left(\frac{3}{2}+\frac{1}{2}|\frac{\tilde{A}}{A}\Theta_2|^2
+\frac{3}{2}|\frac{c_H}{A}v_{24}^2|^2
+\frac{1}{2}|\frac{\tilde{c_H}}{A}\Theta_2|^2\right)=
|v_{\overline{5}}|^2\left(\frac{3}{2}+\frac{1}{2}|
\frac{\tilde{A}}{A}\Theta_1|^2
+\frac{3}{2}|\frac{b_H}{A}v_{24}^2|^2
+\frac{1}{2}|\frac{\tilde{b_H}}{A}\Theta_1|^2\right)
\end{equation}

So, while the minimization conditions discussed above  
fix the value of the product $v_{5}v_{\overline{5}}$, requiring vanishing
D-terms adds eq. (\ref{FEWF})
and fixes the value of
$v_{5}$ and $v_{\overline{5}}$ (and therefore of all the remaining vevs)
as functions of the potential
parameters. 

As a conclusion we have that vacuum alignment in flavor space
$v_i\propto (1,1,1)$, that allows us to
obtain the correct phenomenology in the context of the considered model, arises
in a natural way from the analysis of the
superpotential, under the condition that the VEVs of $\textbf{15}_{\textbf{H}}$ and $\overline{\textbf{15}}_{\textbf{H}}$ 
are neglected in comparison  with the other
scales of the model.

\section{Conclusions}

In this paper we have achieved the possibility to reproduce the nice features of the
$A_{4}$ group, with regard to the mixing of leptons, inside a renormalizable $SU(5)$ theory.
Even if the GUT scale is very close to the Planck scale, in fact, we think that renormalizability has to be a
fundamental characteristic of the considered unification theory, in order to avoid the presence of higher dimensional operators as fundamental
blocks in the construction of the mass matrices and to improve the predictivity of the model.

In our model the neutrino mass matrix comes from the presence of an heavy $SU(2)_{L}$ scalar triplet embedded into the $\textbf{15}_{\textbf{\textbf{H}}}$ representation of $SU(5)$,
while in order to obtain the correct phenomenology at GUT scale we need to introduce an extended Higgs sector, as described in Sec. \ref{Higgses} where the presence of the four dimensional reducible representations of $A_4$ is claimed.
As expected \cite{Picariello:2006sp,Picariello:2007yn} we are not able to
reproduce with the only aid of $A_{4}$ symmetry the hierarchy between the
masses;
 on the contrary the mixing angle
in the CKM-matrix and in the PMNS-matrix, the
 latter being described by the tri-bimaximal
mixing, are reproduced in a very clear way.
With repect to our previos work \cite{Ciafaloni:2009ub}, where a combination
of Type I and Type III Seesaw mechanism was considered for generating
neutrino masses, the model considered here with a Type II mechanism
constitutes an improvement, since no fit is needed in order to generate the
desired tribimaximal mixing. Moreover, as we have shown, minimizing the
potential produces the needed vacuum alignment in a natural
way in a finite region of the potential parameters. 

\section*{Acknowledgments}
Two of us (M.P. and A.U.) would like to thank the organizers and the participants to the ``{\em XLIV$^{th}$ Rencontres de Moriond 2009 - Electroweak Interactions and Unified Theories"} where part of this work have be performed.

\bibliographystyle{../TeX/JHEP}
\bibliography{../SU5_A4/Flavor_models}

\begin{thebibliography}{10}

\bibitem{Altarelli:2009wt}
   For a report on neutrinos masses and mixings, see for instance G.~Altarelli,
   \emph{Status of Neutrino Masses and Mixing in 2009},
   arXiv:0905.3265 [hep-ph] and references therein.

\bibitem{Harrison:2002er}
  P.~F.~Harrison, D.~H.~Perkins and W.~G.~Scott,
 \emph{ Tri-bimaximal mixing and the neutrino oscillation data},
  Phys.\ Lett.\  B {\bf 530}, 167 (2002)
  [arXiv:hep-ph/0202074].

\bibitem{King:2001uz}
  S.~F.~King and G.~G.~Ross,
  \emph{Fermion masses and mixing angles from SU(3) family symmetry},
  Phys.\ Lett.\  B {\bf 520}, 243 (2001)
  [arXiv:hep-ph/0108112].

\bibitem{King:2003rf}
  S.~F.~King and G.~G.~Ross,
  \emph{Fermion masses and mixing angles from SU(3) family symmetry and
  unification},
  Phys.\ Lett.\  B {\bf 574}, 239 (2003)
  [arXiv:hep-ph/0307190].

\bibitem{deMedeirosVarzielas:2005qg}
  I.~de Medeiros Varzielas, S.~F.~King and G.~G.~Ross,
  \emph{Tri-bimaximal neutrino mixing from discrete subgroups of SU(3) and  SO(3)
  family symmetry},
  Phys.\ Lett.\  B {\bf 644}, 153 (2007)
  [arXiv:hep-ph/0512313].


\bibitem{Mohapatra:2006pu}
  R.~N.~Mohapatra, S.~Nasri and H.~B.~Yu,
  \emph{S(3) symmetry and tri-bimaximal mixing},
  Phys.\ Lett.\  B {\bf 639} (2006) 318
  [arXiv:hep-ph/0605020].


\bibitem{Bazzocchi:2008ej}
  F.~Bazzocchi and S.~Morisi,
  \emph{S(4) as a natural flavor symmetry for lepton mixing},
  arXiv:0811.0345 [hep-ph].





\bibitem{Altarelli:2005yp}
  G.~Altarelli and F.~Feruglio,
  \emph{Tri-bimaximal neutrino mixing from discrete symmetry in extra
  dimensions},
    Nucl.\ Phys.\  B {\bf 720} (2005) 64
  [arXiv:hep-ph/0504165].



\bibitem{Hagedorn:2008bc}
  C.~Hagedorn, M.~A.~Schmidt and A.~Y.~Smirnov,
 \emph{ Lepton Mixing and Cancellation of the Dirac Mass Hierarchy in SO(10) GUTs
  with Flavor Symmetries T7 and Sigma(81)},
    Phys.\ Rev.\  D {\bf 79} (2009) 036002
  [arXiv:0811.2955 [hep-ph]].


\bibitem{Feruglio:2007uu}
  F.~Feruglio, C.~Hagedorn, Y.~Lin and L.~Merlo,
 \emph{ Tri-bimaximal neutrino mixing and quark masses from a discrete flavour
  symmetry},
  Nucl.\ Phys.\  B {\bf 775} (2007) 120
  [arXiv:hep-ph/0702194].



\bibitem{Lin:2008aj}
  Y.~Lin,
  \emph{A predictive A4 model, Charged Lepton Hierarchy and Tri-bimaximal Sum
  Rule},
  arXiv:0804.2867 [hep-ph].

\bibitem{Lin:2009bw}
  Y.~Lin,
  \emph{Tri-bimaximal Neutrino Mixing from A(4) and 
$\theta_{13} \sim \theta_C$},
  arXiv:0905.3534 [hep-ph].

\bibitem{King:2009mk}
  S.~F.~King and C.~Luhn,
  \emph{A new family symmetry for SO(10) GUTs},
  Nucl.\ Phys.\  B {\bf 820} (2009) 269
  [arXiv:0905.1686 [hep-ph]].

\bibitem{Picariello:2006sp}
  M.~Picariello,
  \emph{Neutrino CP violating parameters from nontrivial quark-lepton correlation:
  a S(3) x GUT model},
  Int.\ J.\ Mod.\ Phys.\  A {\bf 23} (2008) 4435
  [arXiv:hep-ph/0611189].


\bibitem{Altarelli:2009kr}
   G.~Altarelli and D.~Meloni,
   \emph{A Simplest A4 Model for Tri-Bimaximal Neutrino Mixing},
   J.\ Phys.\ G {\bf 36} (2009) 085005
   [arXiv:0905.0620 [hep-ph]].

\bibitem{Altarelli:2009gn}
   G.~Altarelli, F.~Feruglio and L.~Merlo,
   \emph{Revisiting Bimaximal Neutrino Mixing in a Model with S4 Discrete
   Symmetry},
   JHEP {\bf 0905} (2009) 020
   [arXiv:0903.1940 [hep-ph]].

\bibitem{Feruglio:2008ht}
   F.~Feruglio, C.~Hagedorn, Y.~Lin and L.~Merlo,
   \emph{Lepton Flavour Violation in Models with A4 Flavour Symmetry},
   Nucl.\ Phys.\  B {\bf 809} (2009) 218
   [arXiv:0807.3160 [hep-ph]].



\bibitem{Ding:2008rj}
  G.~J.~Ding,
  \emph{Fermion Mass Hierarchies and Flavor Mixing from $T'$ Symmetry},
  Phys.\ Rev.\  D {\bf 78} (2008) 036011
  [arXiv:0803.2278 [hep-ph]].


\bibitem{Bazzocchi:2007au}
  F.~Bazzocchi, S.~Morisi and M.~Picariello,
  \emph{Embedding A4 into left-right flavor symmetry: Tribimaximal neutrino mixing
  and fermion hierarchy},
  Phys.\ Lett.\  B {\bf 659} (2008) 628
  [arXiv:0710.2928 [hep-ph]].


\bibitem{Cai:2006mf}
  Y.~Cai and H.~B.~Yu,
  \emph{An SO(10) GUT Model with $S4$ Flavor Symmetry},
  Phys.\ Rev.\  D {\bf 74}, 115005 (2006)
  [arXiv:hep-ph/0608022].

\bibitem{Morisi:2007ft}
  S.~Morisi, M.~Picariello and E.~Torrente-Lujan,
 \emph{ A model for fermion masses and lepton mixing in SO(10) x A4},
  Phys.\ Rev.\  D {\bf 75} (2007) 075015
  [arXiv:hep-ph/0702034].

\bibitem{Bazzocchi:2008rz}
  F.~Bazzocchi, S.~Morisi, M.~Picariello and E.~Torrente-Lujan,
 \emph{ Embedding A4 into SU(3)xU(1) flavor symmetry: Large neutrino mixing and
  fermion mass hierarchy in SO(10) GUT},
    J.\ Phys.\ G {\bf 36}, 015002 (2009)
  [arXiv:0802.1693 [hep-ph]].


\bibitem{Altarelli:2008bg}
  G.~Altarelli, F.~Feruglio and C.~Hagedorn,
  \emph{A SUSY SU(5) Grand Unified Model of Tri-Bimaximal Mixing from A4},
  JHEP {\bf 0803}, 052 (2008)
  [arXiv:0802.0090 [hep-ph]].







\bibitem{Ciafaloni:2009ub}
  P.~Ciafaloni, M.~Picariello, E.~Torrente-Lujan and A.~Urbano,
  \emph{Neutrino masses and tribimaximal mixing in Minimal renormalizable SUSY
  SU(5) Grand Unified Model with A4 Flavor symmetry},
  arXiv:0901.2236 [hep-ph].

\bibitem{Dorsner:2007fy}
  I.~Dorsner and I.~Mocioiu,
  \emph{Predictions from type II see-saw mechanism in SU(5)},
  Nucl.\ Phys.\  B {\bf 796}, 123 (2008)
  [arXiv:0708.3332 [hep-ph]].

\bibitem{Altarelli:2005yx}
  G.~Altarelli and F.~Feruglio,
  \emph{Tri-Bimaximal Neutrino Mixing, A4 and the Modular Symmetry},
  Nucl.\ Phys.\  B {\bf 741} (2006) 215
  [arXiv:hep-ph/0512103].











\bibitem{Picariello:2007ss}
  M.~Picariello, B.~C.~Chauhan, J.~Pulido and E.~Torrente-Lujan,
  \emph{Predictions from non trivial Quark-Lepton complementarity},
    Int.\ J.\ Mod.\ Phys.\  A {\bf 22}, 5860 (2008)
  [arXiv:0706.2332 [hep-ph]].




	

\bibitem{Picariello:2007yn}
  M.~Picariello,
  \emph{Predictions for $\mu \rightarrow e\gamma$ in SUSY from non trivial Quark-Lepton
  complementarity},
  Adv.\ High Energy Phys.\  {\bf 2007}, 39676 (2007)
  [arXiv:hep-ph/0703301].










\end{thebibliography}



\providecommand{\href}[2]{#2}\begingroup\raggedright\endgroup

\end{document}